\documentclass[manuscript]{aastex}
\usepackage{amsmath}

\shorttitle{Escaping Particle fluxes in the atmospheres of close-in
exoplanets} \shortauthors{Guo}

\begin{document}


\title{Escaping Particle fluxes in the atmospheres of close-in exoplanets: I. model of hydrogen}


\author{J. H. Guo\altaffilmark{1,2}; guojh@ynao.ac.cn}
\affil{National Astronomical Observatories/Yunnan Observatory,
Chinese Academy of Sciences, P.O. Box 110, Kunming 650011, China}



\altaffiltext{1}{Key Laboratory for the Structure and Evolution of
Celestial Objects, Chinese Academy of Sciences}


\begin{abstract}
A multi-fluid model for an atomic hydrogen-proton mixture in the
upper atmosphere of extrosolar planet is presented when the
continuity and momentum equations of each component have been
already solved with an energy equation. The particle number density,
the temperature distribution and the structure of velocity can be
found by means of the model. We chose two special objects, HD
209458b and HD 189733b, as discussion samples and the conclusion is
that their mass loss rates predicted by the model are in accordance
with those of observation. The most important physical process in
coupling each component is charge exchange which tightly couples
atomic hydrogen with protons. Most of the hydrogen escaping from hot
Jupiters is protons, especially in young star-planet system. We
found that the single-fluid model can describe the escape of
particles when the mass loss rate is higher than a few times
$10^{9}$ g/s while below $10^{9}$ g/s the multi-fluid model is more
suitable for it due to the decoupling of particles. We found that
the predicted mass loss rates of HD 189733b with the assumption of
energy-limit are a factor of 10 larger than that calculated by our
models due to the high ionization degree. For the ionized wind which
is almost compose of protons, the assumption of energy-limit is no
longer effective. We fitted the mass loss rates of the ionized wind
as a function of $F_{UV}$ by calculating the variation of the mass
loss rates with UV fluxes.

\end{abstract}





\section{Introduction}

The discovery that the hot Jupiter HD 209458b is losing mass was
rather unexpected (Recently, Lecavelier des Etangs et al. (2010)
have found atmospheric evaporation in HD 189733b. It is the second
extrasolar planet whose mass loss has been detected.). The excess
absorption in Lyman-alpha first found by Vidal-Madjar et al. (2003) (VM03)
and later confirmed by Linsky et al. (2010) could be explained either by mass loss of
the atmosphere due to XUV energy input from host stars (Lammer et
al. 2003; Lecavelier des Etangs et al. 2004; Yelle 2004, 2006; Tian
et al. 2005; Garcia Munoz 2007; Penz et al. 2008; Murray-clay et al.
2009; Lammer et al. 2009) or by charge exchange between the stellar
wind and the planetary escaping exosphere (Holmstr\"{o}m et al.
2008). For the former, all of these models describe the thermal
particle escape. For the latter, Erkaev et al. (2005) and
Holmstr\"{o}m et al. (2008) brought forward loss of nonthermal
neutral atoms due to interaction between the stellar wind and the
exosphere (more details see Ekenb\"{a}ck et al. 2010). The model of
Erkaev et al. (2005) underestimates evidently the particle loss
rates, but Ekenb\"{a}ck et al. (2010) modeled the production of
neutral hydrogen and match the feature of $Ly\alpha$ absorption. It
is not easy to distinguish which process dominates $Ly\alpha$
absorption more. Ben-Jaffel \& Hosseini (2010) found that either
energetic HI of stellar origin or thermal HI populations in the
planetary atmosphere could fit $Ly\alpha$ observations. Koskinen et
al. (2010)(K10) used an empirical model to analyze UV transit
depths, and their results showed that observations can be explained
solely by absorption in the upper atmosphere while the process of
charge exchange could not be necessary. It is also noteworthy that
the properties of planetary magnetic field are unclear until now.
Thus, it is important for both thermal models and nonthermal models
to calculate self-consistently the deflection distance around planet
by using magnetohydrodynamics models.

The observations of transit in hot jupiters have found that
the mass loss rates are in the range $10^{9}-10^{11}$g/s (VM03; Linsky et
al. 2010; Lecavelier des Etangs et al. 2010). The effect of
evaporation has been used to research the evolution of exoplanet.
Baraffe et al. (2004, 2005) considered the effect of mass loss on
the evolution of exoplanet, and mass loss rates varying from
$10^{-12} M_{J}/year$ to $10^{-8} M_{J}/year$ are obtained from
their calculations. However, an escape rate $10^{2}$ lower
than Baraffe's(2004, 2005) is found by Hubbard et al. (2007a,
2007b). Their results showed that the moderate mass loss rates could be
more appropriate for the mass function of observation. Guo (2010)
further investigated the influence of mass loss on the tidal
evolution of exoplanets, and found that the effect of mass loss on
tidal evolution can not be neglect for those planets with initial
mass $< 1 M_{J}$ and initial orbital distance $<0.1$AU. It has also
suggested that evaporation can lead to a significant modification to
the nature of planet so as to form the planetary remnants
(Lecavelier des Etangs et al. 2004, 2007; Penz \& Micela 2008; Penz et al. 2008; Davis
\& Wheatley 2009).

At present, theoretical models focus on macroscopic material escape
and the photochemistry is incorporated in their models (Yelle 2004;
Tian et al. 2005; Garcia-Munoz 2007; Penz et al. 2008; Murray-Clay
et al. 2009). All of these models suggested that the EUV and X-ray
energy input by host stars is the main energy source for the escape
of atmosphere. However, the escape process of different species may
be dominated by different physical mechanism, and the winds around
exoplanets orbiting their host stars at distant separation could be
tenuous. Thus it must be argued whether the heavier species decouple
from the light elements in tenuous winds (OI and CII transit depths
of 13\% and 7.5\% obtained by Vidal-Madjar et al. (2004) indicated
that both Oxygen and Carbon in the atmosphere of HD 209458b are
lost.). Moreover, a detailed physical model is helpful probably to
distinguish the origin of hydrogen around HD 209458b. Because much
micro-physics has not included into hydrodynamic simulations, the
study to the behaviors of species by using multi-fluid models may be
worthwhile.

With consideration of the process radiative transfer, this paper
aims to calculate planetary atomic hydrogen and proton loss rates
through the solution to their mass, momentum and energy equations. A
great deal of microscopic physics processes are covered in the mass
and momentum equations (Section 2.1). The code is designed for
ordinary equations with one or more critical points (Section 2.2).
We use Henyey method to calculate these equations (Section 2.4).
The results are presented in Section 3. Of especial
interest are what makes atomic Hydrogen and proton decoupling
(Section 4.1)? In Section 4.2 we discuss the properties of ionized
wind and fit the mass loss rate as a function of UV flux. We
summarize my results in Section.5

\section{The model}


The model describes the steady state, radial expansion of plasma
containing three species: atomic hydrogen (h), proton (p) and
electron (e). Each species has its own continuity and momentum
equations and is described by a particle density $n_{s}$ and
velocity $u_{s}$. I do not include $H_{2}$ in this model because the
thermosphere of close-in planet should be composed primarily of H
and $H^{+}$. The location of transition from $H_{2}$ to H is about
at 1.1$R_{P}$ (Yelle 2004). In addition, with the assumption of
$T_{h}=T_{p}=T_{e}$, only one energy equation for electron is given
in the calculations.

\subsection{Equations and assumptions}
As this model deals with a mixture of atomic hydrogen and proton,
the following processes are considered: photoionization,
recombination and charge exchange. The most important collision
between neutral hydrogen and ion is charge exchange. Other
collisions, as hard sphere collision, between neutral and ionized
particle are not important. The continuity and momentum equations
for each species including the source/sink terms which relate to
photoionization/recombination and charge exchange for a species $j$
read as
\begin{equation}
\frac{1}{r^{2}}\frac{d(r^{2}n_{j}u_{j})}{dr}=\frac{\delta
n_{j}}{\delta r} (j=h,p)
\end{equation}

\begin{equation}
u_{h}\frac{du_{h}}{dr}+\frac{1}{m_{h}n_{h}}\frac{dP_{H}}{dr}+\frac{GM_{p}}{r^{2}}-\frac{3GM_{*}r}{a^{3}}=\frac{\delta
M_{s}^{h}}{\delta r}
\end{equation}

\begin{equation}
u_{p}\frac{du_{p}}{dr}+\frac{1}{m_{h}n_{p}}\frac{dP_{p}}{dr}-\frac{eE}{m_{H}}+\frac{GM_{p}}{r^{2}}-\frac{3GM_{*}r}{a^{3}}=\frac{\delta
M_{s}^{p}}{\delta r}
\end{equation}

where G is the gravitational constant, partial pressure
$P_{j}=n_{j}k_{B}T$, temperature $T$, hydrogen atom mass $m_{H}$ and
planetary mass $M_{p}$. The term $\frac{3GM_{*}r}{a^{3}}$ denotes
the tidal gravity. $M_{*}$ and $a$ are stellar mass and semi-major
axis, respectively. $E$ is the charge separation electric field.

Quasi-neutrality and zero-current was assumed in the derivation of
momentum equation, thus, $n_{e}=n_{p}$ and $u_{e}=u_{p}$. Both
inertia and gravity are negligible in the electron momentum equation
because of the small mass of electrons, so we can express the
electric field E as

\begin{equation}
eE=-\frac{1}{n_{e}}\frac{dP_{e}}{dr}.
\end{equation}

The two terms: $\frac{\delta n_{j}}{\delta r}$ and $\frac{\delta
M_{s}^{j}}{\delta r} (j=h,p)$, represent the source and sink terms
caused by elastic and inelastic collisions with other species,
respectively. The sources and sinks for the particle flux density
are due to photoionization and recombination. The term of resonant
charge exchange $H, H^{+}\longleftrightarrow H^{+}, H$ is included
in the momentum equation but not in the continuity equation because
before and after the interaction the same particles are present.
With the assumptions and the definitions as above we write the
collision terms as

\begin{equation}
\frac{\delta n_{h}}{\delta r}=n_{p}\gamma _{rec}-n_{h} \gamma _{pho}
\end{equation}

\begin{equation}
\frac{\delta n_{p}}{\delta r}=n_{h}\gamma _{pho}-n_{p} \gamma _{rec}
\end{equation}

\begin{equation}
\frac{\delta M_{s}^{h}}{\delta
r}=-\gamma_{rce}n_{p}(u_{h}-u_{p})-\gamma_{rec}\frac{n_{p}}{n_{h}}(u_{h}-u_{p})
\end{equation}

\begin{equation}
\frac{\delta M_{s}^{p}}{\delta
r}=-\gamma_{rce}n_{h}(u_{p}-u_{h})-\gamma_{pho}\frac{n_{h}}{n_{p}}(u_{p}-u_{h})
\end{equation}

where $\gamma_{pho}=\frac{F_{UV}e^{-\tau}}{h\nu_{0}}\sigma_{\nu0}$
$s^{-1}$ and
$\gamma_{rec}=2.7\times10^{-13}(\frac{T}{10^{4}})^{-0.9}n_{p}$
$s^{-1}$ (Murray-Clay et al. 2009) represent the photoionization and
recombination rates, respectively.
$\gamma_{rce}=1.12\times10^{-8}(\frac{T}{10^{4}})^{1/2}[1-0.12log(\frac{T}{10^{4}})]^{2}$
$cm^{3}s^{-1}$ is the rate of charge exchange (Geiss \& Burgi 1986).

We have assumed that ions and atoms have the same temperatures,
which is justified by the high collision rates. For the proton, the
temperature equilibration time in atomic hydrogen is
\begin{equation}
t_{eq}=\frac{1}{n_{h}\gamma_{rce}}
\end{equation}
Due to $n_{h} \sim 10^{5}-10^{11} cm^{-3}$, $T \sim 10^{4}$K, we
have $t\sim 10^{-3}-10^{3}$s. Ionization time of about $10^{5}$s
(Yelle 2004) in the upper atmosphere is two orders of magnitude
greater than the temperature equilibration time. Following Schunk
(1975) the energy equation for electrons is given as

\begin{equation}
\frac{3}{2}n_{e}u_{e}k_{B}\frac{dT}{dr}-k_{B}Tu_{e}\frac{dn_{e}}{dr}=\sum_{j}
\sqrt{2}Z_{j}^{2}\nu_{ee}m_{j}(u_{j}-u_{e})^{2}+H-L
\end{equation}

The right terms of the equation denote in sequence elastic
collisions with heavy ions, heating and cooling. As the assumptions
of charge neutrality and zero-current the term of elastic collisions
is zero. The heating process in the mixed flows is complex. The
stellar radiation ionizes the species to produce high energy
photoelectrons which share their energy with other species by
collisions.

A fully description to the process is beyond the scope of this
paper. I follow the model of Murray-Clay et al. (2009) to describe
heating from photoionization, but a coefficient,
$\alpha=\frac{n_{p}}{2n_{p}+n_{h}}$, is used to denote the fraction
which is shared with electrons. Therefore,

\begin{equation}
H=\alpha \varepsilon F_{UV} e^{-\tau}\sigma_{\nu0}n_{h}
\end{equation}

where $\varepsilon=(h\nu_{0}-13.6eV)/h\nu_{0}$ is the fraction of
photon energy deposited as heat, $h\nu_{0}=20$eV, and
\begin{equation}
\tau=\sigma_{\nu0}\int^{\infty}_{r}n_{h} dr
\end{equation}
is the optical depth.

Murray-Clay et al. (2009) pointed out that the main contribution to
cooling is $Ly\alpha$ radiation. Although the radiation is emitted
by the atoms and ions, the excitation of the atoms and ions is due
to electron collisions. Thus, the process is an energy loss
mechanism for the electrons. The cooling is
\begin{equation}
L=7.5\times 10^{-19} n_{h}n_{p}e^{-118348/T}  erg cm^{-3}s^{-1}.
\end{equation}

\subsection{Critical points}
The equations considered here have the similar structure as the
solar wind. From Eqs. (1)-(8) and (10), we find that there exist two
singular (sonic) points where the velocity derivatives can not be
determined by this set of equations. These singularities are
important because the transonic solution is only one that can yield
flows from subsonic velocity at the base of the wind to supersonic
velocity at the outside of the singularity. At these points, the
velocity derivatives must be obtained by additional conditions such
as regularity conditions.

Eliminating all derivatives other than the velocity derivative from
the H atom and proton momentum equations leads to a coupled
differential equations of the momentum, which can be written in the
matrix form
\begin{equation}
\left(
  \begin{array}{cc}
    a_{hh} & a_{hp} \\

    a_{ph} & a_{pp} \\
  \end{array}
\right) \left(
  \begin{array}{c}
    \frac{du_{h}}{dr} \\

    \frac{du_{p}}{dr}  \\
  \end{array}
\right)=\left(
  \begin{array}{c}
    b_{h} \\
   b_{p} \\
  \end{array}
\right)
\end{equation}

where
\begin{equation}
 a_{hh}=(\frac{k_{B}T}{m_{H}u_{h}}-u_{h})
\end{equation}

\begin{equation}
 a_{pp}=(\frac{10}{3}\frac{k_{B}T}{m_{H}u_{p}}-u_{p}),
\end{equation}

where vector $\textbf{b}$ does not contain any of functions of the
derivatives of the variables $n_{j}$, $u_{j}$ (j=h,p),T.

In the case of the solar wind the locus of singularity can be
defined by the points where the determinant of matrix
$D=a_{hh}a_{pp}-a_{ph}a_{hp}$ vanishes (Burgi 1992). In this model
we find $a_{ph}=0$, thus $a_{hh}*a_{pp}=0$ can be forced into
$a_{hh}=0$ and $a_{pp}=0$. It means that there exist two separate
critical points for atomic hydrogen and proton respectively, where
their velocities equal their sound velocities.

\subsection{Boundary conditions}
Six boundary conditions are necessary to solve equations (1)-(3),
(10) and (12). At the bottom of the flow, we set $T_{0}=1000$K. Here
the particle densities are assumed to be $
n_{h}+n_{p}=\rho_{0}/m_{H}$. According to the model of Murray-Clay
et al. (2009) we assumed that the value of density is $4\times
10^{-13} g/cm^{3}$. In the bottom regions the collisions between
atomic hydrogen and proton are frequent, thus we assumed
$u_{h}=u_{p}$. We need additional three boundary conditions to close
the set of equations. Two boundary conditions are provided by the
regularity condition. These conditions can be written as
\begin {equation}
\frac{dH_{i}}{dr}=\sum_{j}(\frac{\partial H_{i}}{\partial
y_{j}})(\frac{\partial y_{j}}{\partial r})=0.
\end {equation}
$H_{i}$ is the momentum equation of species $i$, and $y_{j}$ is for
variables r, T, $n_{k}$, $u_{k}$,(k=h,p). These boundary conditions
are added to corresponding mesh points. However, in calculations we
find that these conditions will result in numerically unstable. When
the total number of mesh points is increased the approximate
conditions $du_{k}/dr=0$ (k=h,p) can be used successfully to
substitute Equation (17)(Nobili \& Turolla 1988). For the last
boundary conditions we assumed the optical depth
$\tau_{c}^{h}=0.0023$ in the sonic points of atomic hydrogen, which
implies that $\tau_{c}^{h}$ equals the optical depth between the
sonic point and Roche lobe (Murray-Clayet et al. 2009).

\subsection{Numerical method}
The momentum equations have two singularities at the two points
$u_{p}^{2}=\frac{10/3\kappa T}{m_{H}}$ and $u_{h}^{2}=\frac{\kappa
T}{m_{H}}$ (Eqs. 15 and 16) for proton and atomic hydrogen,
respectively. The explicit form of equations (2) and (3) can be
written as

\begin{equation}
y^{'}+\frac{g(x,y)}{f(x,y)}=0
\end{equation}

where $f(x,y)$ vanishes in the sonic points. Normally, the velocity
of flow increases with radius. At a specific point (sonic point)
$f(x,y)$ begins to change sign. Thus the critical points can be
found at the location where the sign of $f(x,y)$ changes. To solving
boundary problems, Henyey method is convenient and the critical
points appearing within the integration domain can be treated
consistently (Nobili \& Turolla 1988). As a modified Newton-Raphson
method, the method includes the critical conditions into the set of
linearized flow equations. All the variables $Y_{i}$ at half-mesh
point are interpolated as
\begin{equation}
Y_{x_{i+1/2}}=\frac{Y_{i+1}+Y_{i}}{2}
\end{equation}

while derivatives are approximated by finite differences
\begin{equation}
Y'_{x_{i+1/2}}=\frac{Y_{i+1}-Y_{i}}{x_{i+1}-x_{i}}.
\end{equation}

In our calculations we allow for the specification of one condition
every time a critical point is met.

\section{Results}

For the comparison with single-fluid model, firstly, The
Murry-Clay et al. (2009) model is solved by their method and
subsequently did again by Henyey method. No significant difference
is found by different solving methods, while some minor differences
could exist due to detailed numerical configuration. An advantage of
using Henyey method is to obtain the solutions once at whole domains
and to self-consistently handled the several critical points in
numerical configuration.

\subsection{HD 209458b}
In this section we applied our model to a typical and particular
planet sample: HD 209458b with the radius of 1.4 $R_{J}$, the mass
of 0.7 $M_{J}$ and the semi-major axis of 0.05 (Murray-Clay et al.
2009). The upper limit on the observed orbital eccentricity is
0.028(Ibgui \& Burrows 2009), so we assumed a circular orbit in the
calculations. The mass of host star is 1$M_{\odot}$.

The results for HD 209458b are shown and the single-fluid results of
Murray-Clay et al. (2009) are also plotted in Fig.1. Seen from Fig.1
clearly, except for a detailed difference, the multi-fluid model
predicts much the same trends of temperature, velocity and particle
number density as the single-fluid model does. The number density of
atomic hydrogen decreases (upper left of Fig. 1) and the ionization
fraction increases with radius (lower right of Fig. 1). The fact
that as much as 80\% of hydrogen at $R=5R _{p}$ is ionized is a
consequence of being irradiated by host star. With the increase of
radius the optical depth becomes smaller, even near to zero, thus
the radiation from star can freely penetrate the wind and ionize
particles. Current studies show the wind temperature can attain
8000-10000K at about 2$R_{p}$ (Yelle 2004; Tian et al. 2005; Garcia Munoz 2007; Penz et al. 2008;
Murray-Clay et al. 2009) and our results also verify the fact. The
cooling by radiation and PdV work make the temperature
reduced from 8000K and decrease to about 5000K in the upper
atmosphere. The inversion of temperature in the upper atmosphere is
a typical feature of planetary atmosphere. To compare our results
with the results of other hydrodynamic models (Yelle 2004; Tian et al. 2005;
Garcia Munoz 2007; Penz et al. 2008 and Murray-Clay et al. 2009),
we found that all models obtain similar temperature profiles besides the model
of Tian et al. (2005) which predicts a monotonously increasing temperature profile
and a higher velocity structure. A possible reason is that Tian et al.(2005) applied
a higher radiative heating. 

The transonic wind driven by stellar XUV has two sonic points: one is for atomic hydrogen and
the other is for protons. Because protons bear an external electric
force, E, with the assumption of the same temperature in both
particles, the sonic point extends to outside of the wind. The
velocity of H and $H^{+}$ is almost same. When one particle travels
outward, the other particle drags it inward. The collision between
two particles leads to a tight couple for their kinetics.

We have described the photoionization, recombination and charge
exchange rates in Section 2.1. Evidently, charge exchange dominates
the other two collision process ($n\gamma_{rce} \gg \gamma_{pho},
\gamma_{rec}$). To show the effect of charge exchange we calculate a
model without charge exchange (In fact, we encounter numerical
problems when the charge exchange is fully neglected. Thus we remain
the term and multiply the term by a factor of $10^{-3}$.). Figure 2
displays that protons decouple from atomic hydrogen. If the most
important process is neglected, the behavior of each particle looks
like a single ``fluid".

\begin{figure*}
\epsscale{.80} \plotone{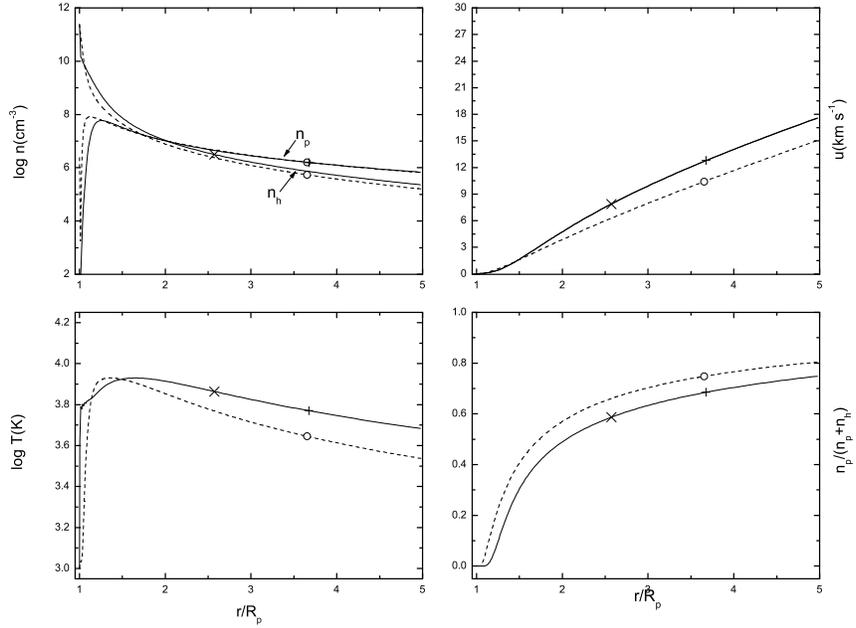}  \caption{The wind model of HD
209458b. Dashed lines are the results of Murray-Clay et al. (2009).
Solid lines represent our results. On each panel, the sonic point of
single-fluid is marked with circles. The sonic points of $H$ and
$H^{+}$ are marked with $\times$ and $+$, respectively. Number
densities (upper left), velocities (upper right), temperatures
(lower left) and the ionization fraction (lower right) are plotted
as functions of altitude. Note that, for the number densities, the
solutions of single-fluid and multi-fluid are indistinguishable. }
\end{figure*}

Note that the ratios of two velocity components decline with the
increase of radius but they rise at a point again. The phenomenon
can be explained by the behavior of photoionization and
recombination rates. At the inner of the wind, the amount of
momentum transferred by the process of photoionization and
recombination is not enough, therefore neutral hydrogen decouple
from protons. With the decrease of optical depth and temperature,
the effect of photoionization and recombination rates on
transferring momentum increases at the outer of the wind so that the
value of $u_{h}/u_{p}$ also increases. A similar phenomenon will be
discussed in Section 4.1.

\begin{figure*}
\epsscale{.80} \plotone{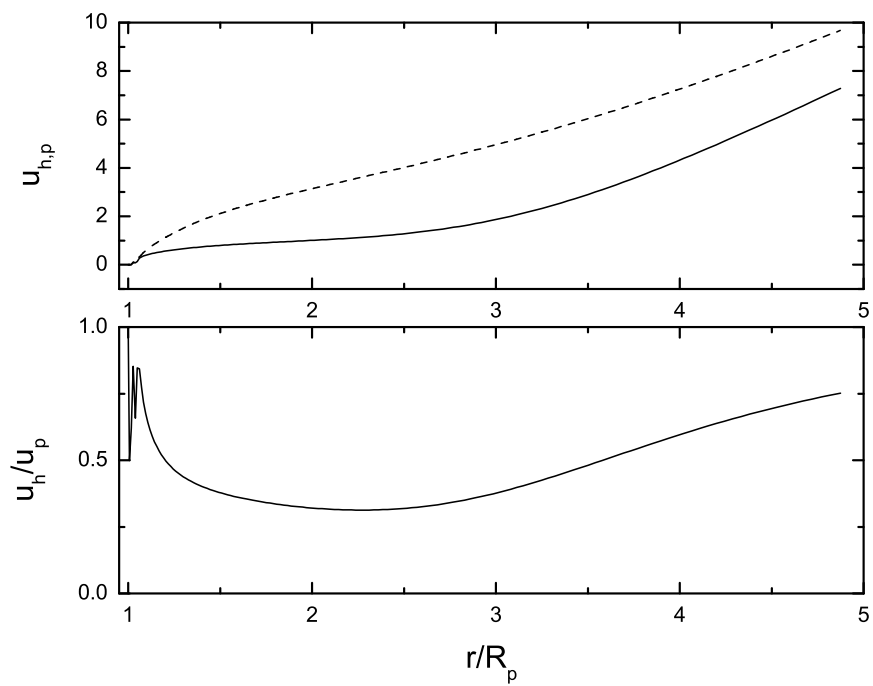}  \caption{ The effect of charge
exchange on velocity profiles. The velocities of neutral hydrogen
(solid line) and protons (dashed line) are indicated in upper panel.
Lower panel shows the ratio of velocity, $u_{h}/u_{p}$, when the
effect of charge exchange is neglected.}
\end{figure*}

The mass loss rates are defined as

\begin{equation}
\dot{M}=4\pi
m_{H}((r_{c}^{h})^{2}n_{h}u_{h}+(r_{c}^{p})^{2}n_{p}u_{p})
\end{equation}

where $r_{c}^{h}$ and $r_{c}^{p}$ are the sonic points of atomic
hydrogen and protons respectively, $n$ and $u$ are corresponding
number density and velocity. The mass loss rate of $\dot{M}=9\times
10^{10}$ g/s obtained by applying the solution over the entire
planetary surface is higher than that of $\dot{M}=3\times 10^{10}$
g/s obtained by Murray-Clay et al. (2009), but both of them accord
approximately with that of VM03 in observation. To fit the observed
profile of $Ly\alpha$, K10 assumed that the mean temperature of
thermosphere is $T=8000-10000$K and the upper boundary is located at
$R \sim 2.9R_{p}$ where the number density is $n=2.6\times 10^{7}
cm^{-3}$. Our calculation results show that the particle number
density at the radius ($R \sim 2.9R_{p}$) is $n=5\times 10^{6}
cm^{-3}$. A significant difference between their model and ours is
that the assumption of hydrostatic equilibrium in their models. Our
results show that the sonic points of $H$ and $H^{+}$ are at
2.57$R_{p}$ and 3.67$R_{p}$, respectively. It means that the
assumption of hydrostatic equilibrium is not unacceptable in the
region range from 1$R_{p}$ to 2.57$R_{p}$. However, the number
density distribution could be different due to the difference in
physical detail. For example, our results that about 60\% hydrogen
is ionized at 2.9$R_{p}$ do not support the assumption of K10 that
the atmosphere is mostly ionized above 2.9$R_{p}$. In addition, in
the assumption of hydrostatic equilibrium the profile of number
density is flatter than that of hydrodynamics, which can lead to
high optic depth in the wings of line.

It is convenient to define the mass loss rates of neutral hydrogen
and protons as
\begin{equation}
\dot{M}_{h,p}=4\pi m_{H}((r_{c}^{h,p})^{2}n_{h,p}u_{h,p}).
\end{equation}
Our results indicate that the mass loss rates of neutral hydrogen
and protons are $3.4\times 10^{10}$g/s and $5.6\times 10^{10}$g/s,
respectively.

To fit the observations of HD 209458b, two scenarios can supply a satisfactory fit.
In the first case, thermal hydrogen atoms are enough to be used to fit the lyman alpha transit profile
so that the energetic atoms are not necessary. In the second case, superthermal (hot)
hydrogen atoms are required in order to fit the observations if depleted
thermal hydrogen atoms are assumed (Ben-Jaffet et al. 2010). Superthermal hydrogen
atoms in the atmosphere can be formed via the
absorption of stellar UV radiation(Shematovich 2010). As hydrogen
atoms have an excess of kinetic energy, they could be locally
thermalized with the surrounding particles if the density of
atmosphere is enough high. But, at the upper atmosphere,
superthermal hydrogen atoms may escape from the atmosphere due to
their excess of kinetic energy. A mass loss rate of $3.4\times10^{9}$ g/s lower
than the observational value for HD 209458b has been
estimated by Shematovich (2010), and their results
shown that those superthermal
hydrogen atoms have velocities which are about 20 km $s^{-1}$. Therefore,
it is unlikely that superthmrmal hydrogen atoms can explain the observed
velocities at the wing of line.
Note that the location of transition from $H_{2}$ to H is about 1.1$R_{p}$(Yelle 2004;
also refer Shematovich 2010), this hints that there superthermal hydrogen atoms
can not directly escape from the atmosphere of planet, but they must exchange
energy and momentum with cool background particles in large-scale ranges. Exactly,
the influence of superthremal hydrogen should be included into future hydrodynamic model.

\subsection{HD 189733b}
HD 189733b is the second extrasolar planet whose atmospheric
evaporation has been detected (Lecavelier des Etangs et al. 2010).
According to observations, they constrain the escape rate of atomic
hydrogen to be between $10^{9}$ to $10^{11}$ g/s and EUV flux is
10-40 times the solar value, namely, $F_{EUV}$ $\simeq$
$1-4\times10^{4}$ erg/$cm^{2}$/s. The value may be changed to
$2.5-10\times10^{4}$ erg/$cm^{2}$/s if the emission of H I
$Ly\alpha$ is also included. The observed value has been validated
by X-ray observations. The luminosities of HD 209458 and HD 189733
in X-ray are measured as $\log L_{x}=26.12$ and $\log L_{x}=28.18$
erg/s (Sanz-Forcada et al. 2010) by XMM-Newton. Because the orbital
distance of HD 189733 is 1.5 times closer to its host star than that
of HD 209458b, the planet receives X-ray radiation which is 300
times more than that of HD 209458b from host star. If assume that
the flux in X-ray is proportional to UV flux, HD 189733b receives a
UV flux of $F_{HD 19733b}=F_{HD 209458b}\times 300 \sim 10^{5}
erg/cm^{-2}/s$.

The planet has a mass $M_{p}=1.13 M_{J}$, radius $R_{p}=1.16 R_{J}$
and the semi-major axis $a=0.03$AU (Bakos et al. 2006; Winn et al.
2007; Southworth 2010). The mass of its host star is $0.8 M_{\odot}$
(Nordstr\"{o}m et al. 2004). Comparing with HD 209458b, the mass of
HD189733b is 1.7 times larger than that of HD 209458b, and its
radius is smaller. This means that the potential well of HD 189733b
is about twice deeper than that of HD 209458b. Thus, the effect of
strong X-ray flux can be balanced to a certain extent by its lager
potential well.

\begin{figure*}
\epsscale{.80} \plotone{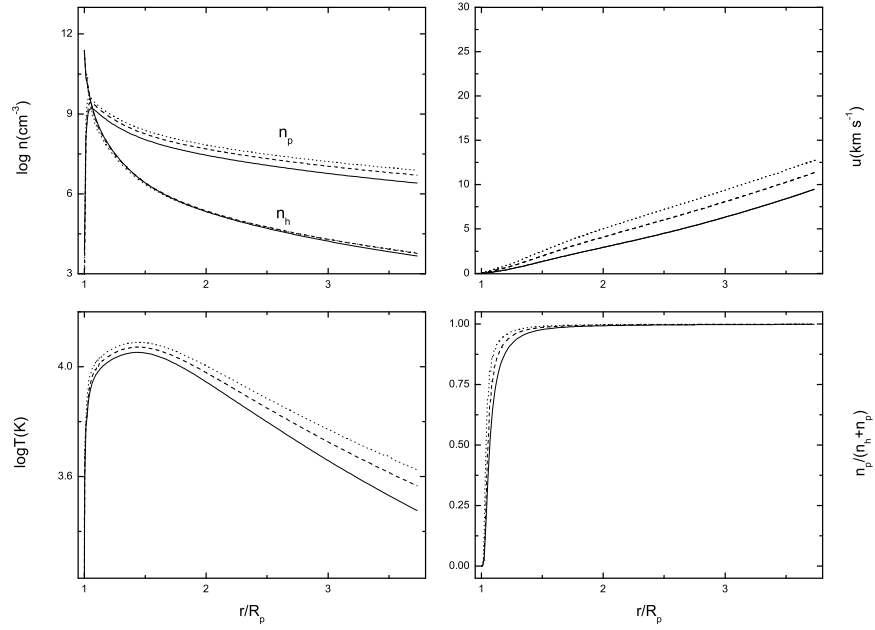}  \caption{The wind model of HD
189733b. From top to bottom, the values of UV flux are $10^{5}$
(dotted lines), $5\times 10^{4}$ (dashed lines) and $5\times 10^{4}$
erg/$cm^{2}$/s (solid lines). On each panel, number densities (upper
left), velocities (upper right), temperatures (lower left) and the
ionization fraction (lower right) are plotted as functions of
altitude. }
\end{figure*}

Figure 3 displays the results for HD 189733b with different values
of $F_{UV}$. The mass loss rates are $4.8 \times 10^{10}$, $1.1
\times 10^{11}$ and $1.98 \times 10^{11}$g/s for $F_{UV}$= $2\times
10^{4}$, $5\times 10^{4}$ and $10^{5}$ erg/$cm^{2}$/s. The mass loss
rate is sensitive to UV flux received from its host star. Lecavelier
des Etangs et al. (2010) found that the escape rate between $10^{9}$
to $10^{11}$ g/s can fit Lyman $\alpha$ absorption of observation.
However, we find that a value of $10^{9}$ g/s can be obtained only
when the UV flux decreases to $10^{3} erg/cm^{2}/s$ which is one
order of magnitude smaller than the lower limit of observed value
(1$\sigma$ level). If the value of $F_{UV}$ is $10^{4}$
erg/$cm^{2}$/s, the mass loss rate of HD 189733 is $2.4\times
10^{10}$ g/s.

In contrast to the results of HD 209458b, as much as 80\% of
hydrogen at $R=1.2R _{p}$ is ionized and the wind is almost fully
ionized outside $R=1.6R _{p}$. The mass loss rate of neutral
hydrogen is only of the order of magnitude of $10^{8}$ g/s due to
the photoionization of strong UV irradiation form host star (The
number density of $H$ is of the order of magnitude of $10^{6}
cm^{-3}$ at $R=1.6R _{p}$ for the model of $F_{UV}= 5\times 10^{4}$
erg/$cm^{2}$/s. The corresponding temperature at the radius is
11000k.). Actually, the wind is almost composed of protons. Under
this circumstance, we wonder whether the neutral hydrogen can
produce adequate absorption that can be detected (Due to the steep
decline of the number density of neutral hydrogen, the optical depth
of wing of line can be very low.). If the amount of atomic hydrogen
is not adequate to fit the observations, the fact that the transits
of HD 189733b in HI Lyman-$\alpha$ have been observed may imply that
others mechanisms can play an important role. A direct comparison
with the model of Lecavelier et al. (2010) can answer the question.
Unfortunately, the physical details of the model of Lecavelier et
al.(2010) were not published.

\section{Discussions}
\subsection{Decoupling of species}
We have modeled the particle escape by a multi-fluid model. For the
planet HD 209458b we do not find significant differences between the
results of multi-fluid model with that of single-fluid model. In
most cases the very close-in planets are bathed by strong XUV
radiation from their host stars. The species of gas are tightly
coupled by collisions so that the description of single-fluid is
accurate. But, under certain special circumstances the description
of single-fluid should be revisited or substituted, for example,
when the planetary wind is tenuous or the irradiation from star is
weak. For high number density, the frictional force is able to
transfer sufficient momentum from one species to other species.
However, decrease in the number density may lead to lowering in the
frictional force, i.e. decoupling is possible.

The case is modeled by the method mentioned in this paper in order
to discuss the possibility of the decoupling for $H$ and $H^{+}$.
Figure 4 shows the ratios of two velocity components for HD 209458b
and HD 189733b in their standard models. Neutral and ionized
hydrogen have the same velocity over whole region. In order to test,
HD 209458b and HD 189733b are moved to two times their original
separations, namely, $a=0.1$ and 0.06AU, and other parameters are
retained. Seen from Figure 4, both the particles remain the same
velocity profile. We also note that the number densities of
particles only lower a factor of 2 than that of standard model.

\begin{figure*}
\epsscale{.80} \plotone{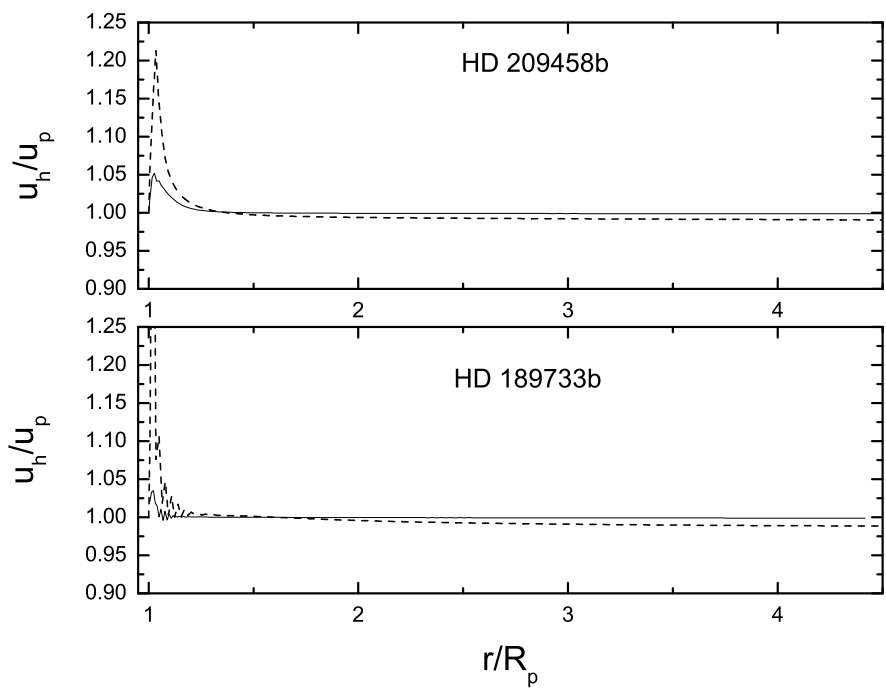}  \caption{Plot of $u_{h}/u_{p}$ for
HD 209458b and HD 189733b (dashed line: two times original
separation; solid line: original separation.). In order to test the
assumption of single-fluid, HD 209458b and HD 189733b are moved to
two times their original separations. No significant diffusion was
found in both cases. }
\end{figure*}

The conclusion that even at large separation the description of
single-fluid is still real seems to be drawn. However, it must be
noticeable that the star HD 189733 is younger than HD 209458 (The
ages of HD 189733 and HD 209458b are about 1.15Gyr (Sanz-Forcada et
al. 2010) and 4Gyr (Guo 2010), respectively.), which means more XUV
radiation can be emitted by HD 189733. If the age of HD 189733 is
same as that of HD 209458, the conclusion may be different. For
testing such a theoretical hypothesis, we calculated many cases with
different values of $F_{UV}$ for HD 189733b. Seen from Figure 5,
there exists a critical mass loss rate below which decoupling can
occur (We assumed that decoupling would occur when $u_{h}/u_{p}$ is
smaller than 95\% in the calculations.). The results show that a
significant diffusion between neutral hydrogen and protons occurs
while the mass loss rate is below $1.3\times 10^{9}$ g/s. Therefore
the critical value is about $10^{9}$ g/s.

The conclusion is also verified by testing other cases. In order to
decrease the mass loss rate of HD 209458b, we artificially increased
the mass of the planet as 0.85$M_{p}$ and decreased the density of
lower boundary by a factor of 10. The results show that ionized
hydrogen has a higher velocity profile and a clear decoupling occurs
throughout the wind. Finally, we obtained a mass loss rate of
$\dot{M}=5\times10^{8}$ g/s.

\begin{figure*}
\epsscale{.80} \plotone{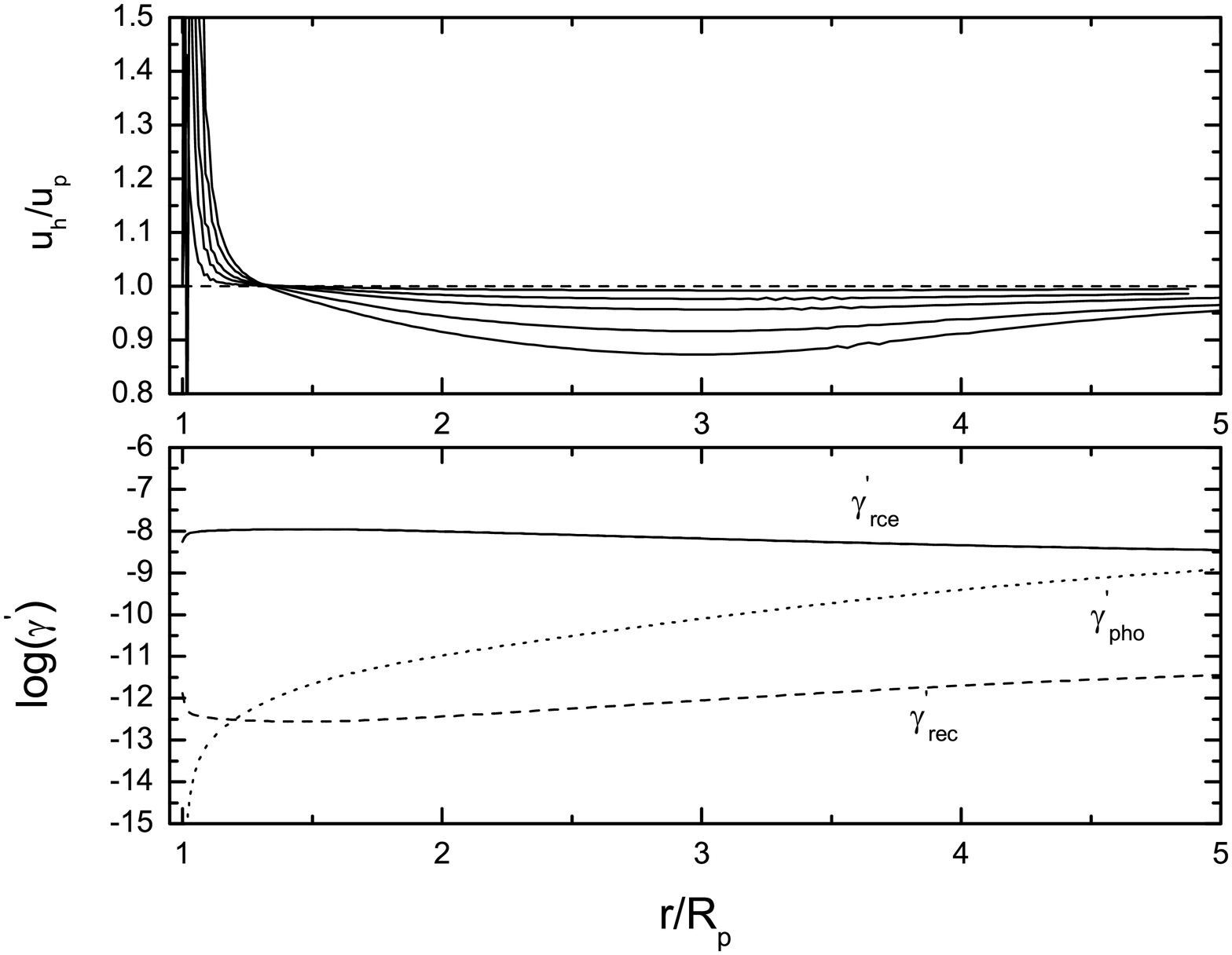}  \caption{The velocity ratios of
atomic hydrogen to protons (top panel) and the photoionization,
recombination and charge exchange rates (bottom panel). In the top
panel, the mass loss rates are $8.8\times 10^{8}$, $1.3\times
10^{9}$, $2.4\times 10^{9}$ g/s, $4.2\times 10^{9}$ and $5.9\times
10^{9}$. The corresponding UV fluxes are $450$, $630$, $1080$,
$1800$ and $2500$ erg/$cm^{2}$/s (from bottom to top). The bottom
panel only shows the photoionization, recombination and charge
exchange rates in the model with $F_{UV}=450$ erg/$cm^{2}$/s. For
comparison, we defined these rates as:
$\gamma_{pho}'=\gamma_{pho}/n_{p}$,
$\gamma_{rec}'=\gamma_{rec}/n_{p}$ and $\gamma_{rce}'=\gamma_{rce}$
$cm^{3}s^{-1}$.}
\end{figure*}

In order to obtain a common flow, the momentum gained by one species
should be shared by other species. This requires that the
characteristic time scale for slowing down the fast particle by
interaction with the other slow particle should be smaller than the
time scale of flow. The time scale of flow can be estimated as
$R_{s}/u$, where $R_{s}\sim 0.1R_{J}$ and u in $10^{5}-10^{6}$ cm.
The slowing down time scale, $t_{s}$, is approximately as
$1/(n_{h}\gamma_{rce})$. The condition of decoupling can be
estimated by equaling two time scale. Therefore the critical number
density of neutral hydrogen is about $10^{4}-10^{5}$ $cm^{-3}$. Due
to the $n_{p}/n_{h}\sim 10$, the critical mass loss rate is at the
order of magnitude of $10^{9}$g/s.


As mentioned in Section 3.1, the ratios of two velocity components
behave as parabolic curves. To maintain the common flow, a
sufficient amount of momentum must be transferred between two
components. The most important process of transferring momentum is
charge exchange. Near the bottom of the wind, the wind is relatively
dense so that the process of momentum exchange is effective.
However, with the increase of radius the rate of charge exchange
decreases (The solid line at the bottom of Figure 5.). Thus, the
transfer of momentum is also decrease with radius. We note that the
process of photoionization and recombination can also redistribute
momentum from one component to the other component, and the
variations of photoionization and recombination rates (dashed and
dotted lines) are contrary to that of charge exchange rate.
Therefore, at the outer of the wind the momentum transfered by
photoionization and recombination also plays a role. Finally, there
is minimum value of $u_{h}/u_{p}$ in the middle of wind.

\subsection{Dependence of $\dot{M}$ on UV fluxes}
Lammer et al. (2003) presented that the energy deposition of X-ray
and UV radiation from parent star can lead to a high temperature,
and that a hydrodynamic process can occur in planetary atmosphere.
The mass loss rates of energy deposition tightly depend on the
fluxes of XUV radiation. In general young stars can radiate more
energy than old ones in XUV band. The energy-limit mass loss rate
can be written as
\begin{equation}
\dot{M}=\frac{3\eta\beta^{3}F_{s}}{G\rho K(\xi) }
\end{equation}
where $\beta$ is the ratio of the expansion radius R1 to the
planetary radius $R_{p}$, R1 is altitude where the XUV radiation is
absorbed, $R_{p}$ represents the distance form the center of planet
to the 1 bar pressure level in the atmosphere, $\eta$ is heating
efficiency, and $\rho$ is the mean density of planet. Here
$K(\xi)=1-\frac{3}{2\xi}+\frac{1}{2\xi^{2}} < 1$ is a non-liner
potential energy reduction factor due to the stellar tidal force
(Erkaev et al. 2007), and $\xi=d(\frac{4\pi\rho}{9M_{*}})^{1/3}$ is
Roche lobe boundary distance, where $M_{*}$ is the mass of star and
$d$ is the orbital distance.

Based on the hydrodynamic model of Watason et al. (1981),
Lammer et al. (2003) estimated $\beta=3$, but it could be unit
according to recent hydrodynamic models which showed that the
expansion radius could be 1 - 1.5 $R_{p}$ (Yelle 2004; Murray-Clay
et al. 2009). Lammers et al. (2009) also thought the mass loss rate
was overestimated with $\beta=3$ by Barraffe et al. (2004). With the
full energy-limited condition, the heating efficiency $\eta=100\%$.
In fact, the heating efficiency is about 25\%. In this paper, we set
$\beta=1.1$ and heating efficiency $\eta=0.1-0.25$ (Murray-Clay et
al. 2009 and refer therein). Thus, Equation (23) describes a
modified energy-limit approach.

For comparison with the energy-limited mass loss rate, we calculated
the mass loss rate of HD 189733b as a function of UV flux and the
results is shown in Figure 6 (left panel). The mass loss
rates given in our models are a factor of 3-10 lower than those
calculated by Equation (23) at the assumption of $k(\xi)=1$ and
$\eta=0.1-0.25$. For completeness we also calculated the
single-fluid model (Murray-Clay et al. 2009) and found a systemic
difference in comparing with the mass loss rates predicted by the
model of this paper. For single-fluid model, it can predict a
comparable value for $\dot{M}$ when the heating efficiency in the
energy-limit method is decreased to $\eta=0.1$ (the left panel of
Fig.6). However, this is not consistent with our results. The mass
loss rates calculated by single-fluid model is still higher than
these of our model. With the increase of $F_{UV}$, the ratios of
$\dot{M}_{single}/\dot{M}_{this paper}$ decrease from
5($F_{UV}$=450) to 2.5($F_{UV}=10^{5}$). As discussed in Section
3.2, our results fit the observations well. Thus, a lower heating
efficiency is required for high ionized wind.

Our calculation results show that the winds are highly ionized and
almost composed of protons even if the UV fluxes are assumed at a
low level. The high ionization degree can be explained as the
consequence of low mass loss rate. Even if the UV flux is at same
level, the mass loss rate of HD 189733b should be lower than that of
HD 209458b due to the larger potential well of HD 189733b. Thus, for
HD 189733b, the low mass loss leads to the low optical depth and
high ionization degree. Given the energy equation, it is clear that
the photoionization heating is proportional to the number density of
neutral hydrogen. With the assumption of energy-limit most of the UV
radiation energy is deposited as heat (due to the low ionization
degree), which is used to lift material out of the gravitational
potential well. Thus, the condition of energy-limit results in
higher mass loss rates. In the case of ionized wind the material is
mainly composed of protons. Only a little of UV radiation can be
transformed as heat, and further goes into PdV work. Murray-Clay et
al. (2009) found that at high $F_{UV}$ the flow is
radiation-recombination-limited (at low flux $\dot{M} \propto
F_{UV}^{0.9}$; at high flux $\dot{M} \propto F_{UV}^{0.6}$), and an
almost isothermal wind is predicted. In contrast to the case of
radiation-recombination-limit, our results show a "normal"
temperature profile. It hints that heating is balanced by PdV work
rather than radiation cooling. Thus, we
can summarize that the modified energy-limit approach can be used in
the case of low or moderate ionization degree, but is unsuccessful
for the high ionization winds. The conclusion can be validated by
comparison with HD 209458b, we find that the mass loss rate
calculated by Equation (23) can predict a reasonable observation
value for HD 209458b.


\begin{figure}
\begin{minipage}[t]{0.5\linewidth}
\centering
\includegraphics[width=2.9in,height=2.2in]{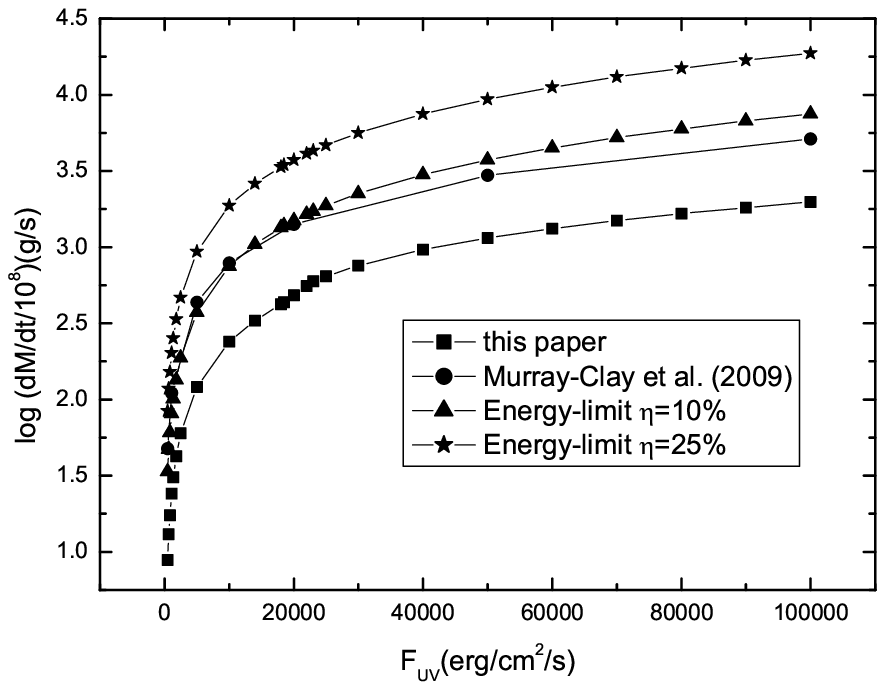}
\end{minipage}
\begin{minipage}[t]{0.5\linewidth}
\centering
\includegraphics[width=2.9in,height=2.2in]{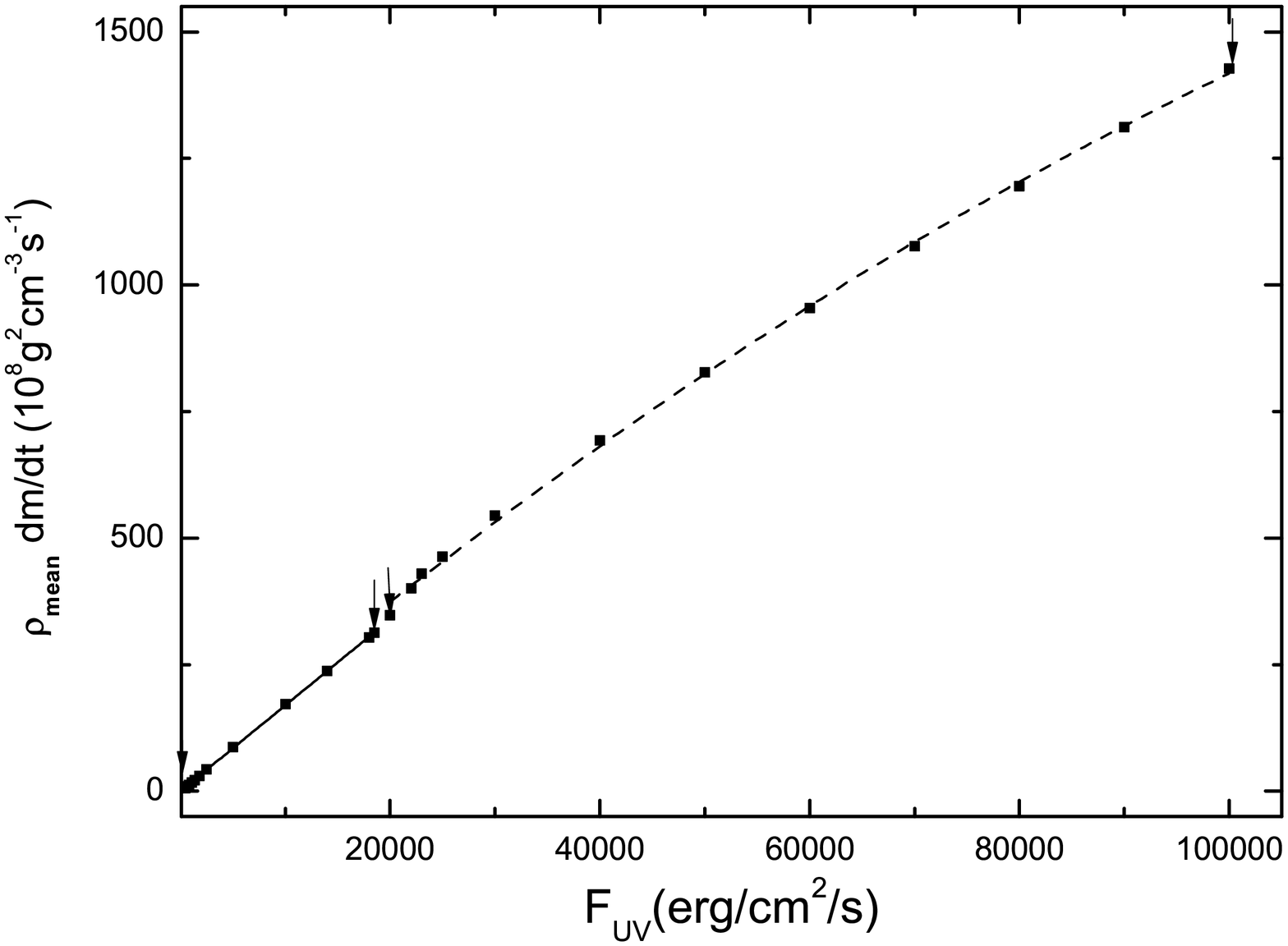}
\end{minipage}
\caption{Left:The behaviors of mass loss rate as a function of UV fluxes.
The results of energy-limit($\blacktriangle$:heating efficiency $\eta=10\%$; $\bigstar$: heating efficiency $\eta=25\%$.)
clearly show larger mass loss rates than those predicted
by hydrodynamic models($\bullet$: the model of Murray-Clay et al. 2009; $\blacksquare$: the model of this paper.).
Right:The mass loss rates as a
function of UV fluxes. For $F_{UV}\leqslant 2000$ erg/$cm^{2}$/s, we
used a liner-fitting (solid line). We fitted the rest of the data
with a polynomial (dashed line).}
\end{figure}

The interesting results motivated us to fit the mass loss rate as a
function of UV flux in the case of ionized wind. Seen from Figure 6,
it is clear that the straight line can depict the lower part, and
the upper part can be fitted by using a polynomial (a jump appears
at $F_{UV}=2000$). Finally, the mass loss rate can be express as


\begin{equation} \label{eq:2}
\dot{M}\bar{\rho}= \left\{
\begin{array}{ll}
-0.24+0.017F_{UV}, F_{UV}\leqslant 2\times10^{4} \\
34.1+0.018F_{UV}-3.92\times10^{-8}F_{UV}^{2}, 10^{5}\geqslant
F_{UV}>2\times10^{4}
\end{array}
\right.
\end{equation}
where$\bar{\rho}$ is the mean density of planet.

Note that Equation (24) is only appropriate for ionized planetary
wind. But, it is not an easy task to determine whether the wind is
ionized. To maintain an ionized wind, the photoionization rate
should be larger than recombination rate, namely, $\gamma_{pho}
> \gamma_{rec}$. Thus, we have
\begin{equation}
\frac{F_{UV}}{h\nu_{0}}n_{h} e^{-\tau}\sigma_{\nu_{0}}
> n_{p}^{2} \gamma_{rec}.
\end{equation}

At $R=1.5R_{p}$, the optical depth $\tau \sim 0$, and $T\approx
10000$k. If the wind is ionized highly at $R=1.5R_{p}$, the value
of$n_{p}/n_{h}$ is about 10-100. Inequality (25) can be changed to

\begin{equation}
\frac{F_{UV}}{h\nu_{0}}\frac{\sigma_{\nu_{0}}}{\gamma_{rec}}
> \frac{n_{p}}{n_{h}}n_{p}.
\end{equation}

With the assumption of $F_{UV}=10^{3}$erg/$cm^{2}$/s, the left of
inequality (26) is about in the order of magnitude $10^{9}$. Using
the hydrostatic density profile, we can estimate the particle number
density at $R=1.5R_{p}$. For HD 209458b, $n_{p}\sim 7\times10^{9}$

so that $\frac{n_{p}}{n_{h}}n_{p}$ is of the order of $10^{11}$.
Because the left term of inequality (26) is smaller than the right
term, the wind of HD 209458b is not ionized. For HD 189733,
$n_{p}\sim 10^{8}$. Thus the inequality (26) can be fulfilled
roughly if $F_{UV}\geqslant 10^{3}$erg/s.

\subsection{The effect of magnetic field}
Many planets in our solar system have magnetic fields of their own.
For example, the magnetic field at the surface of Jupiter is about 4.3G.
What role the magnetic field plays on the upper atmosphere depends on
their field strength. For the planets without intrinsic magnetic fields,
the magnetic fields could be induced at the interaction area between the
stellar and planetary winds. However, the occurrence of strong
magnetic fields occurring at the surfaces of planets can lead to significant
changes in the upper atmospheres of planets. Trammell et al.(2011)
have presented a 3D isothermal magnetohydrodynamic model based on the
stellar wind model(Mestel 1968) although some physical details and
the interaction between stellar and planetary winds are not considered into it.
To include fully physical processes, for example, magnetic field and
collision of winds, a powerful 3D MHD
model is required. It is beyond the scope of this paper. However, the potential role
of magnetic field can be discussed in the current 1D model. In
the following discussion, we assume the planetary flows are
spherically symmetric, namely, the magnetic field does not change the
geometry of planetary wind within magnetosphere, but the boundary of
magnetosphere constraints the range of planetary flows.

The most important effect of magnetic field on the hydrodynamic
escape is whether the magnetic field can change the locations of
sonic points for different species. If the boundary of magnetosphere
is greater than that of sonic point, the planet can still emit a
transonic wind. Otherwise, the planetary wind will be suppressed to
subsonic flow, even quenched. Grie$\beta$meier et al. (2004) have
estimated the stand off distance of magnetosphere, $R_{s}$, by
pressure balance and found that the stand-off distance of HD 209458b
are about 2.6-3.8 $R_{p}$ which varied with different
rotation velocities. Our results show that the locations of sonic
points of H and $H^{+}$ are comparable with the distance of stand-off. As
discussed previously(Section 3.1), the flow of atomic hydrogen could
be transonic, but the escape of proton could be suppressed to
subsonic flow. However, seen from Equations (2),(3),(7) and (8),
the drag force of H by $H^{+}$ is negative if the velocity of H
is higher than that of $H^{+}$. Thus the subsonic behavior of $H^{+}$ could
result in the decrease of the velocity of H. In the process, charge
exchange shifts momentum from one species with high velocity to the other
species with low velocity. One possibility for the case is that both
the flows become subsonic within magnetosphere and attain a new
equilibrium status. As a consequence of the case, the mass loss
rates of H and $H^{+}$ could decrease a factor of a few(Murray-Clay et
al. 2009). In fact, the stand-off distance depends strongly on the
strength of magnetic field, and the field strength could vary with the mass
and evolution of planet(Reiners \& Christensen 2010; Scharf 2010).
The capture to radio signal can answer what the intrinsic magnetic field is,
however, no signal from the
observations has been found so far (Lazio et al. 2010).

Note that the pressure of gas is omitted in the calculation of Grie$\beta$meier et al. (2004)
so that the stand-off distance could increase if
the gas contributes a significant pressure fraction. Johansson et
al. (2009) indicated the influence of expanding atmosphere on the
stellar interaction. For unmagnetized exoplanets, the ram pressure
of ionosphere can push the bow shock toward the host stars. The
results hints that if the contribution of gas to the total pressure is
included, the magnetopause of exoplanet could be outside of sonic
points.

The ionization degree of hydrogen is not controlled explicitly by its velocity structure. Equation(1) shows clearly that
the ionization of hydrogen is dominated by photoionizaiton. The magnetic fields have
an indirectly effect on the ionization by changing
the mass loss rates of planets. If the mass loss rates decrease a factor of a few, we
can predict that there is a higher ionization degree in the atmosphere
because the optical depth will also decrease.

For the exoplanet with an intrinsic dipole magnetic field,
the ions near the equator could be inhibited by planetary
magnetic field while the neutrals can escape freely from the
atmosphere. At the portion of the planetary surface occupied by the closed magnetic fields
hydrostatic equilibrium has been attained. Ions can not escape from the closed
magnetic field although the neutrals can do so. Trammell et
al.(2011) predicted a mass loss rate of $10^{9}$ g/s when the
magnetic field was considered into their model. For the details, reader may refer to the
corresponding 2D or 3D models(Trammell et al.2011; Adams 2011).
Our model should be suitable in the polar regions where the magnetic
lines are open and almost aligned with the radial direction. Thus, the
neutrals can escape from the whole surface of planet. But, the escaping
surface of ions could decrease at least
a factor of 2 (Yelle 2004). For the low ionization wind, the intrinsic dipole magnetic
fields only confine a small part of the wind. For the high ionization wind
in which ions are main composition, the intrinsic magnetic field can
severely decrease the mass loss rate.

\section{Conclusions}
We have developed a multi-fluid model to describe the particle
escape from the upper atmosphere of close-in planet. The continuity
and momentum equations for each component were solved together with
an energy equation. Detailed micro-physical process, for example,
photoionization, recombination and charge exchange, have been
included in our models. Based on Henyey method, the code can treat
systems of ordinary differential equations with one, or more,
critical points.

We calculated two samples of close-in planet, HD 209458b and HD
189733b. Our calculations show that the mass loss rates of two
planets are of the order of magnitude of $10^{10}-10^{11}$ g
$s^{-1}$ which agree with the observed mass loss rates of
$10^{10}-10^{11}$ g $s^{-1}$. By detailed test and verification, we
found that the most important physical process in the atmosphere of
hot Jupiter is charge exchange which tightly couples atomic hydrogen
with protons. Most of the hydrogen escaping from hot Jupiters is
protons, especially in young star-planet system. Thus, the transit
of Lyman-$\alpha$ in HD 189733b could be induced by other unknown
physics processes.

In the assumption of single-fluid all species must remain collide
frequently. Otherwise, the description of single-fluid can not be
used. Our method can be further apply to other case, for example,
the tenuous wind. It is found that decoupling may occur if the mass
loss rates are lower than $10^{9}$ g/s.

We also found that the assumption of energy-limit is not appropriate
for ionized winds. Our model predicted a mass loss rate lower than
that of energy-limit. By calculating the variations of mass loss
rate with UV fluxes we fitted the mass loss rates of ionized wind as
a function of $F_{UV}$.

\section*{Acknowledgments}

This work was supported by National Natural Science Foundation of
China(Nos.10803018)and Western Light Talent Culture Project of The
Chinese Academy of Sciences(08AXB31001). I thank Ruth A Murray-Clay
for helpful discussion in their single-fluid model. I also thank the
referee for feedback that improved the paper.

\clearpage

\end{document}